\documentclass[structabstract]{aa}  
\usepackage{graphicx}
\usepackage{epsfig}
\usepackage{txfonts}
\usepackage{natbib}
\usepackage{lscape}
\bibpunct{(}{)}{;}{a}{}{,} 

\def\asec{\ifmmode ^{\prime\prime}\else$^{\prime\prime}$\fi}

\def\msun{\hbox{M$_{\odot}$}}

\def\degs{\ifmmode ^{\circ}\else$^{\circ}$\fi}
\def\amin{\ifmmode ^{\prime}\else$^{\prime}$\fi}
\def\asec{\ifmmode ^{\prime\prime}\else$^{\prime\prime}$\fi}

\def\degs{\ifmmode ^{\circ}\else$^{\circ}$\fi}
\def\amin{\ifmmode ^{\prime}\else$^{\prime}$\fi}
\def\EE#1{\times 10^{#1}}

\def\cm{\mbox{\,cm}}

\def\cm3{\mbox{\,cm$^{-3}$}}

\def\ergs{\mbox{\,erg~s$^{-1}$}}
\def\ergcmcmcm{\mbox{\,erg~cm$^{-3}$}}

\def\lsim{\!\!\!\phantom{\le}\smash{\buildrel{}\over
 {\lower2.5dd\hbox{$\buildrel{\lower2dd\hbox{$\displaystyle<$}}\over
                                 \sim$}}}\,\,}
\def\gsim{\!\!\!\phantom{\ge}\smash{\buildrel{}\over
{\lower2.5dd\hbox{$\buildrel{\lower2dd\hbox{$\displaystyle>$}}\over
                               \sim$}}}\,\,}

\begin{document}
\title{The nuclear starburst in Arp~299-A: From the 5.0 GHz VLBI radio light-curves to its core-collapse supernova rate}

\author{M. Bondi \inst{1}\thanks{These authors have contributed equally to the 
paper.}
      \and 
      M.A. P\'erez-Torres \inst{2}$^\star$
      \and 
      R. Herrero-Illana \inst{2}
      \and 
     A. Alberdi \inst{2}
}
          
\institute{ Istituto di Radioastronomia di Bologna - INAF, Via P. Gobetti, 101 
40129 Bologna, Italy\\
             \email{mbondi@ira.inaf.it}
         \and
             Instituto de Astrof\'{\i}sica de Andaluc\'{\i}a - CSIC,
             PO Box 3004, 18008 Granada,  Spain 
         }

\date{}

\abstract
{The nuclear region of the Luminous Infra-red Galaxy
 Arp~299-A hosts a recent ($\simeq 10$ Myr), intense burst of 
 massive star formation which is expected to lead to 
numerous core-collapse supernovae (CCSNe). Previous VLBI observations,
carried out with the EVN at 5.0 GHz and with the VLBA at 2.3 and 8.4 GHz,
resulted in the detection of a large number of compact, bright, non-thermal
sources in a region $\lsim$150 pc in size. 
}
{We aim at establishing the nature of all non-thermal, compact
  components in Arp 299-A, as well as estimating its core-collapse 
supernova rate. 
 While the majority of them are expected to be young radio supernovae 
  (RSNe) and supernova remnants (SNRs), a definitive classification is still 
  unclear. Yet, this is very relevant for eventually establishing the 
  core-collapse supernova rate, as well as the star formation rate, 
  for this galaxy. } 
{We use multi-epoch European VLBI Network (EVN) observations taken at
  5.0 GHz to image with milliarcsecond resolution the compact radio sources 
  in the nuclear region of Arp 299-A. We also use one single-epoch 5.0 GHz 
  Multi-Element Radio Linked Interferometer Network (MERLIN)
  observation to image the extended emission in which the
  compact radio sources --traced by our EVN observations-- are embedded.}
{We present the first 5.0 GHz radio light-curve (spanning $\sim$2.5
  yr) of all compact components in the nuclear starburst of Arp~299-A.
  Twenty-six compact sources are detected, 8 of them are new objects
  not previously detected. The properties of all detected objects
    are consistent with them being a  mixed population of CCSNe and SNRs.
  We find clear evidence for at least two new CCSNe, implying a lower limit 
  to the CCSN rate of  $\nu_{\rm SN}\gsim$0.80 SN/yr indicating that the 
 bulk of the current star formation in Arp~299-A is taking place in the 
 innermost $\sim 150$ pc.
{  A few more objects show variability consistent with being recently exploded 
  SNe, but only forecoming new observations will clarify this point.
  Our MERLIN observations trace a region of diffuse, extended emission
  which is cospatial to the region where all compact sources are found
  in our EVN observations. From this diffuse, non-thermal radio
  emission traced by MERLIN we obtain an 
  independent estimate for the core-collapse supernova rate, which is in the
  range $\nu_{\rm SN}=0.40$ - $0.65$ SN/yr, in}
  agreement with previous estimates and our direct estimate of the
  CCSN rate from the compact radio emission. }
{Our $\sim$2.5 yr monitoring of Arp~299-A has allowed us to obtain
  for the first time a direct estimate of the CCSN rate of 
  $\nu_{\rm SN}\gsim 0.80$ SN/yr for the innermost $\sim$150 pc of
  Arp~299-A. }

\keywords{Galaxies: starbursts -- luminosity function, mass function
  -- individual: Arp~299-A -- Stars: supernovae: general -- Radiation
  mechanisms: non-thermal -- Radio continuum: stars}

\titlerunning{The nuclear starburst in  Arp~299-A}

\maketitle
%

\section{Introduction}

According to the $\Lambda$ cold dark matter hierarchical models of galaxy 
formation, more massive galaxies are assembled from smaller ones in a series 
of minor or major merger events \citep[e.g.][]{TT72}.  Nearby examples of 
merger galaxies are often associated with the class of Luminous, LIRG with 
$ 10^{11} L_\odot \le L_{\rm IR} < 10^{12} L_\odot$, and Ultra Luminous Infrared 
Galaxies, ULIRG with $L_{\rm IR} \ge 10^{12} L_\odot$ \citep{Soif89,SM96}.  
Galaxy interactions and mergers can produce stellar bars and/or 
enhanced spiral structure that drive ample quantities of molecular gas and 
dust into the circumnuclear regions of (U)LIRGs, triggering massive star 
formation and possibly accretion onto a supermassive black hole
\citep[e.g.][]{dimatteo07}

Coupling high sensitivity with a wide range of angular resolutions, and being
unaffected by dust obscuration,
radio observations provide an important and unique tool to investigate the 
nuclear components of (U)LIRGs. 
In particular, Very Long Baseline Interferometry (VLBI) observations allowed 
to resolve the nuclear component of local (U)LIRG,
detecting tens of compact objects in the merger nuclei of Arp~220 and
Arp~299 \citep{Smith98, Rovi03, Lons06, Parra07, Neff04, PT09, Ulve09,PT10}.
The non-thermal origin of these compact sources is confirmed by their
high brightness temperatures, implying that most of them are
young core-collapse (radio) SNe (CCSNe) or supernova remnants (SNRs).
Radio spectral index and flux density variability can be used to
distinguish between the two classes of sources \citep{Parra07}, but the results
are usually not conclusive for all sources, and
therefore the number of observed CCSNe is uncertain. 
This is an important parameter, as the observed rate at which massive stars
($M \gsim 8 M_\odot$) 
explode as CCSNe can be used as a direct measurement
of the current star formation rate (SFR) in galaxies, and provides unique 
information about the initial mass function (IMF) of massive stars.
 
The subject of this paper is Arp~299-A, the eastern component 
of the peculiar interacting system, Arp~299,  in an early merging state
\citep{Keel95}.
Arp~299 is at a distance of 44.8 Mpc, 
for a resdhift of $z=0.010811$ as corrected to the Reference Frame defined by 
the 3K Microwave Background Radiation \citep{fixsen96}, and assuming
$H_0 = 73$~km~s$^{-1}$~Mpc$^{-1}$, $\Omega_{\rm matter} = 0.27$ and 
$\Omega_{\rm vacuum}= 0.73$. At the assumed distance, 1\arcsec\ corresponds 
to 217~pc.

Arp~299 has an infrared luminosity $L_{\rm
  IR} \approx 6.7\times 10^{11} L_{\odot}$ \citep{sanders03}, which is
approaching the ULIRG category, and makes of Arp 299 the most
luminous infra-red galaxy in the nearby 50~Mpc. 
HST and NICMOS observations have demonstrated that Arp~299-A is the most
dust-enshrouded source in the system \citep{AH00} and is responsible
for the largest fraction of the far-infrared luminosity \citep{CSG02}. 
Numerous H~II regions populate
the system near star-forming regions, suggesting that star formation
has been occurring at a high rate for the last $\sim$10 Myr
\citep{AH00}.  

Arp 299 hosts recent and
intense star forming activity, as indicated by the relatively high
frequency of supernovae discovered at optical and near-infrared
  wavelengths in their outer, much less extinguished regions
\citep{forti93,treffers93,vanburen94,li98,yamaoka98,qiu99,mattila05,NPO10,
MK10}.
Yet, the innermost $\sim 150$ pc nuclear region of Arp~299-A is heavily 
dust-enshrouded, thus making the detection of SNe very challenging even at 
near-infrared wavelenghts.

Since optical and near-infrared
observations are likely to miss a significant fraction of CCSNe in the
innermost regions of Arp 299-A due to large values of extinction [$A_V
\sim 34-40$ \citep{gallais04,AH09}] and the lack of the
necessary angular resolution, radio observations of Arp 299-A at high
angular resolution and high sensitivity are the only way of detecting new
CCSNe and measuring directly and independently of models its CCSN and
star formation rates.  
In fact, VLBI observations
carried out  since 2003 resulted in the
detection of several compact sources \citep{Neff04,Ulve09,PT09,PT10}, 
one of which (A0) was unambiguously identified as a young CCSN.

In this paper, we present the first results from a high-sensitivity
VLBI monitoring campaign at 5.0 GHz on Arp~299-A. The main goals of
our project are to provide a census of all the compact sources in the 
nuclear region of Arp~299-A, to directly detect new CCSNe and thus 
determine its CCSN rate, and to use 
radio light-curves obtained through multi-epoch observations to help
classifying the sources as CCSNe, SNRs or other classes (e.g. transitional
objects, microquasars).

\begin{table*}
\centering                          
\caption{5.0 GHz (e)EVN observations of Arp 299}             
\label{tab,evn}      
\renewcommand{\footnoterule}{}
\begin{tabular}{lllllll}
\hline \hline
 Date & Experiment  & $\nu$  & Antenna Array & Integration &Resolution &
 r.m.s.$^{\mathrm{b}}$ \\
           & Code               & [GHz]          &  &[hr] & [mas]  & [$\mu$Jy/beam] \\
\hline 
08 Apr 2008 & RP009   & 4.99      &  eEVN (Cm, Jb, Mc, On, Tr, Wb) & 8.2 &7.3 x
6.3 & 40.5 \\
05 Dec 2008 &  RP014A & 4.97      &  eEVN (Cm, Ef, Jb, Kn, Mc, On, Sh,Tr,
Wb) & 6.0 &8.6 x 8.4 &  29.4 \\
12 Jun 2009 & EP063B & 4.99 & EVN (Cm, Ef, Jb2, Mc, Nt, On, Sh, Tr, Ur, Wb,
Ys )&3.5  & 6.9 x 5.1 & 28.5 \\
11 Dec 2009 & RP014C & 5.00 & eEVN  (Ef, Jb$^{\mathrm{a}}$,  Mc, On,Sh,Tr,
Wb, Ys) & 3.4 &10.0 x 7.4 & 37.6 \\
28 May 2010 & EP068A &  4.99 & EVN (Ef, Mc, On, Tr, Ur, Wb, Ys) & 3.5& 7.9 x
6.7 & 32.0 \\
24 Nov 2010  & RP014D & 4.99 & eEVN  (Cm, Ef, Jb, Mc, On, Tr, Ws, Ys, Sh )&3.6& 
10.1 x 6.0 & 29.1 \\
\hline
\end{tabular}   
\begin{list}{}{}
\item[$^{\mathrm{a}}$] Because of a network problem in the UK, Jb
  could only join in the very last hour.
\item[$^{\mathrm{b}}$] The r.m.s. refers to the images restored with a
$10.0\times 8.0$ mas beam used in the analysis.  
\end{list}
\label{tab,evn6cm}
\end{table*}

\section{5.0 GHz EVN observations and data analysis}
\label{sec,observations}

We observed Arp 299-A with the
European VLBI Network (EVN),  which yields milliarcsecond angular
resolution, at a frequency of 5.0 GHz for six epochs, taken at
approximately six-month intervals between April 2008 and December
2010.  Additionally, at two epochs, June 2009 and June 2010, we also used 
the EVN at 1.6 GHz close to two similar observing runs at 5.0 GHz, so as to
obtain spectral index information to aid classifying the nature of the
compact radio sources. The 1.6 GHz images and spectral index  analysis
will be presented in a forecoming paper (P\'erez-Torres et al., in preparation).

Table \ref{tab,evn} shows the log for our observations, including the
observing date, experiment code, observing frequency, array used, time
on-source, 
angular resolution and r.m.s. noise level attained. 
Four epochs have been carried out using the real time eEVN.
The only notable difference between standard disk based EVN  
and real time eEVN observations is that in the latter the data stream is 
not recorded on disk and later correlated,
but is directly transferred to the correlator through fiber links
and processed in real time.
The first two epochs have already been presented and discussed 
in a previous publication \citep{PT09}, but in this paper we take 
advantage of all the six epochs to produce new images and a complete 
analysis. 

All the observing epochs were carried out as phase-referenced 
observations. The first two epochs (RP009 and RP014A)  used a data
recording rate of 512~Mbps with two-bit sampling, for a total bandwidth
of 64~MHz, while the remaining epochs used a data recording rate of 1024~Mbps for a total bandwidth of 128 MHz. 
The data were correlated at the EVN MkIV Data Processor at 
JIVE using an averaging time of 1~s. The telescope systems recorded both 
right-hand and left-hand circular polarization (RCP and LCP) which, after 
correlation, were combined to obtain the total intensity images analyzed in 
this paper.  Usually, 4.5 minute scans of our target source, Arp 299-A, were 
alternated with 1 minute (2 minute for experiment RP009) scans of our phase 
reference source, J1128+5925. The bright sources 3C84,
3C138, 4C39.25 and 3C286 were used as fringe finders and band-pass
calibrators, depending on the epoch.

\subsection{Data calibration and imaging}
\label{sec,cal}

We analyzed the correlated data for each epoch using the NRAO
Astronomical Image Processing System ({\it AIPS};
http://www.aips.nrao.edu).  The visibility amplitudes were calibrated
using the system temperature and gain information provided for each
telescope. Standard channel-based inspection and editing of the data were done
within AIPS.  The bandpasses were corrected using the bright
calibrator 4C39.25, 3C345 or J1128+5925 depending on the experiment.  
We applied standard corrections to the phases of
the sources in our experiment, including ionosphere corrections (using
total electron content measurements publicly available).

The instrumental phase and delay offsets among the baseband converters in each 
antenna were corrected using the phase calibration determined from 
observations of 4C\,39.25, 3C345 or J1128+5925, depending on the experiment.  
The data for the calibrator J1128+5925 were then fringe-fitted in a standard 
manner.  We then exported the J1128+5925 data into the Caltech imaging program 
DIFMAP \citep{shepherd95} for mapping purposes.  We thus determined gain 
correction factors for each antenna. J1128+5925 is a strongly variable source 
with peak-to-through amplitudes exceeding 20\% \citep{Gaba09} and a point-like 
structure at the scales sampled by our observations ($\simeq 10$ mas).  
After this procedure was completed within DIFMAP, the data were read back into 
AIPS, where the gain corrections determined by DIFMAP were applied to the data.

The final source model obtained for J1128+5925 was then included as an
input model in a new fringe-fitting search for J1128+5925, thus
removing the structural phase contribution to the solutions of the
delay and fringe rate for our target source, Arp 299-A, prior to
obtaining the final images.  The
phases, delays, and delay-rates determined for J1128+5925 were then
interpolated and applied to the source Arp 299-A. This procedure
allowed us to obtain the maximum possible accuracy in the positions
reported for the compact components in Figure \ref{fig,mean}. 

Particular care was used in order to image all six epochs in an
homogeneous way. After several tests with different weighting functions, we 
chose to use a natural weighting scheme (except for experiment EP068A
where we used the 'NV' weighting function), stopping the cleaning process
 when the 
absolute value of the peak in the residual image was about 3 times the r.m.s.
and then restoring the images with a beam of $10.0 \times 8.0$ mas,
corresponding to $2.2 \times 1.7$ pc.

\begin{figure*} \centering
\includegraphics[width=120mm,angle=0]{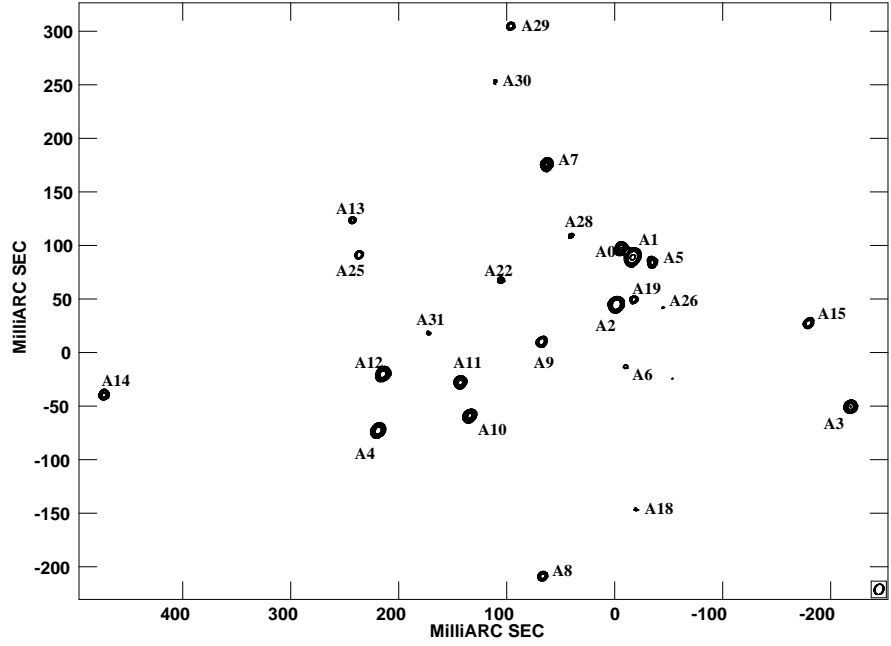}
\caption{EVN image at 5.0 GHz of Arp299-A, obtained by stacking all
six epochs in the image plane. The beam is shown in the bottom-right corner and
the resolution is $10\times 8$ mas P.A. 
$-25^\circ$. The peak is 907 $\mu$Jy/beam, and the first contour is drawn at 
5 times the r.m.s $\sigma$ of 18.5 $\mu$Jy/beam. 
Labeled components, from A0 to A25, correspond to sources already reported 
in \citet{PT09}, from A26 to A31 to new sources at 5.0 GHz. 
Sources A1 and A5 are 
components of the low-luminosity AGN identified by \citet{PT10}}
\label{fig,mean}
\end{figure*}

We point out that we kept the averaging integration time to 1\,s and used
a maximum channel bandwidth in the imaging process of 8\,MHz or 16\,MHz,
depending on the epoch, which results in a maximum degradation of the peak 
response for the component furthest away from the phase center (about 500 mas) 
of less than 2\% and prevents artificial smearing of the images 
\citep[e.g.][]{BS99}.

No self-calibration of the phases, nor of the amplitudes, was
performed on the data, since the peaks of emission were too faint for
such procedures to be applied.

\begin{table*}
\centering                          
\caption{Source list}             
\label{tab,sources1}      
\renewcommand{\footnoterule}{}
\begin{tabular}{lllrcccllll}
\hline \hline
 Name  & RA Off. &  Dec Off.& $<S>$ & $<L>$ & $S_{\rm max}$ & $L_{\rm max}$ &
Other\\
       &    [mas]  &  [mas]     &[$\mu$Jy]& $[\times 10^{26}$ erg s$^{-1}$ Hz$^{-1}$] &
[$\mu$Jy]&[$\times 10^{26}$ erg s$^{-1}$ Hz$^{-1}$]& Name \\
\hline 
33.5941+46.560& $-219$ & $-50$&  380  & 9.1 &   421   & 10.1 & A3\\
33.5991+46.637& $-179$ & $+27$&  249  & 6.0 &   301   &  7.2 & A15\\
33.6163+46.652& $-43$  & $+43$ &  98  & 2.4 &   202   &  4.9 & A26\\
33.6176+46.694& $-35$  & $+84$&  218  & 5.2 &   248   &  6.0 & A5$^{\mathrm {a}}$\\  
33.6195+46.463& $-20$  & $-150$& 105  & 2.5 &   155   &  3.7 & A18\\
33.6198+46.660& $-19$  & $+49$&  159  & 3.8 &   273   &  6.6 & A19\\ 
33.6199+46.699& $-16$  & $+89$&  907  &21.8 &   985   & 23.7 & A1$^{\mathrm {a}}$\\  
33.6207+46.597& $-10$  & $-13$ & 107  & 2.6 &   203   &  4.9 & A6\\ 
33.6212+46.707& $-6$   & $+97$&  399  & 9.6 &   465   & 11.2 & A0\\
33.6218+46.655& $ -1$  & $+45$&  719  &17.3 &   834   & 20.0 & A2\\ 
33.6221+46.692& $+1$   & $+83$ &      &     &   251   &  6.0 & A27\\    
33.6272+46.719& $+44$  & $+109$& 124  & 3.0 &   173   &  4.2 & A28\\ 
33.6301+46.786& $+64$  & $+176$& 415  &10.0 &   501   & 12.0 & A7\\ 
33.6305+46.401& $+67$  & $-210$& 245  & 5.9 &   294   &  7.1 & A8\\ 
33.6307+46.620& $+68$  & $+11$&  269  & 6.5 &   333   &  8.0 & A9\\
33.6344+46.915& $+98$  & $+304$& 163  & 3.9 &   210   &  5.0 & A29\\
33.6355+46.677& $+105$ & $+68$&  149  & 3.6 &   235   &  5.6 & A22\\ 
33.6362+46.863& $+111$ & $+253$& 107  & 2.6 &   140   &  3.4 & A30\\ 
33.6392+46.551& $+135$ & $-59$&  438  &10.5 &   530   & 12.7 & A10\\
33.6403+46.582& $+143$ & $-28$&  338  & 8.1 &   425   & 10.2 & A11\\
33.6441+46.629& $+173$ & $+19$ & 110  & 2.6 &   176   &  4.2 & A31\\
33.6495+46.590& $+215$ & $-20$&  510  &12.2 &   579   & 13.9 & A12\\
33.6500+46.537& $+220$ & $-73$&  541  &13.0 &   576   & 13.8 & A4\\
33.6523+46.701& $+238$ & $+91$&  181  & 4.3 &   277   &  6.7 & A25\\ 
33.6531+46.733& $+243$ & $+124$& 153  & 3.7 &   248   &  6.0 & A13\\
33.6825+46.570& $+474$ & $-40$&  222  & 5.3 &   289   &  6.9 & A14\\ 
\hline
\end{tabular}   
\begin{list}{}{}
\item[$^{\mathrm {a}}$] A1 and A5 are components of the low-luminosity AGN
identified by \citet{PT10}.
\end{list}
\end{table*} 


\begin{table*}
\centering                          
\caption{Source monitoring}             
\label{tab,sources2}      
\renewcommand{\footnoterule}{}
\begin{tabular}{lllllllllll}
\hline \hline
 Name  &  RP009 & RP014A & EP063B & RP014C & EP068A &  RP014D & Label & Var.Flag$^{\mathrm {a}}$ \\
       & [$\mu$Jy]&[$\mu$Jy]&[$\mu$Jy]&[$\mu$Jy]&[$\mu$Jy]&[$\mu$Jy]&    &        &\\
\hline 
33.5941+46.560&   397  &  297   &  382   &  406   &  421   &  386 & A3 &NV\\
33.5991+46.637&  (173) &  301   &  269   &  287   &  274   &  217 & A15& P\\
33.6163+46.652&  (147) & (112)  &$<85$   &$<113$   &  202   &  151 & A26 &\\
33.6176+46.694&   248  &  178   &  215   &  247   &  235   &  207 & A5$^{\mathrm {b}}$ &NV\\
33.6195+46.463&  (139) & 155    &$<85$   & (123)  & (135)  & (121)& A18 &\\
33.6198+46.660&  (122) &  182   &  169   &  273   &  170   &  137 & A19& P\\
33.6199+46.699&   863  &  844   &  917   &  874   &  939   &  985 & A1$^{\mathrm {b}}$ & P\\
33.6207+46.597&   203  &$<88$   & (120)  &$<113$  & (153)  & (122)& A6 & \\
33.6212+46.707&   332  &  396   &  459   &  465   &  380   &  395 & A0 & V\\
33.6218+46.655&   723  &  653   &  675   &  737   &  679   &  834 & A2 & V\\
33.6221+46.692&$<121$  &$<88$   &$<85$   &  251   &$<96$   &$<87$ & A27 &\\
33.6272+46.719&  (152) & (94)  &  179   & $<113$  &  173   & (113)& A28 &\\  
33.6301+46.786&   501  &  382   &  332   &  317   &  484   &  466 & A7 & V\\
33.6305+46.401&   231  &  269   &  189   &  235   &  259   &  294 & A8 &NV\\
33.6307+46.620&   284  &  322   &  333   &  186   &  291   &  253 & A9 & P\\
33.6344+46.915&  (187) & (111)  & (164)  &  211   &  185   &  167 & A29 &NV\\ 
33.6355+46.677&  (174) &  235   & (151)  & (118)  & (106)  &  176 & A22&NV\\
33.6363+46.864&  (129) & (112)  &$<85$   & (140)  & (107)  & 136  & A30& \\
33.6392+46.551&   530  &  424   &  435   &  399   &  423   &  424 & A10& P\\
33.6403+46.582&   301  &  350   &  425   &  386   &  353   &  285 & A11& V\\
33.6441+46.629& $<121$  & (109)  &  176   & (159)  & (145)  &  (89)  & A31 &\\ 
33.6495+46.590&   429  &  579   &  516   &  423   &  518   &  576 & A12& V\\
33.6500+46.537&   535  &  576   &  545   &  543   &  480   &  550 & A4 &NV\\
33.6523+46.701&  (157) &  223   &  277   & (165)  & (167)  &  162 & A25& V\\
33.6531+46.733&   248  & (123)  &  199   & (152)  & (147)  & (98)& A13& V\\
33.6825+46.570&   241  &  221   &  219   & (157)  &  225   &  289 & A14&NV\\
\hline
\end{tabular}   
\begin{list}{}{}
\item[$^{\mathrm{a}}$] V, NV and P stand for variable, non-variable and probably
variable (see discussion in Section~\ref{sec,var} for details).
\item[$^{\mathrm {b}}$] A1 and A5 are components of the low-luminosity AGN
identified by \citet{PT10}.
\end{list}
\end{table*}

\subsection{Source detection, identification, and flux density extraction}
\label{sec,detection}
Our first goal was to identify the {\it bona-fide} sources in the
Arp~299-A field. While at least 16 sources are systematically detected in 
each single epoch above 5 times the r.m.s. (see Table 1), spurious sources 
close to that threshold could appear in any given single epoch. 
For that reason, a 
much more robust way of detecting {\it bona-fide} sources is by combining 
all six epochs, which significantly decreases the noise with respect to an 
individual single-epoch image. The data combination can be done in the 
{\it u-v} plane, by merging the six data-sets
and then producing an image from the resulting data-set, or in the image
plane, by stacking the images produced at each single epoch, to obtain an 
average image.
We chose the latter method (but see \citet{Ulve09} for a different
approach) because in this way we avoid deconvolution problems 
with variable sources in the merged data-set. In addition,  we also get rid of
the spurious sources due to spikes of noise that could be
above the cutoff threshold at any individual epoch (those spikes disappear 
when the images are stacked, since they occur at random, different locations, 
from epoch to epoch).

Figure~\ref{fig,mean} shows the six-epoch 5.0 GHz stacked (averaged) 
image of Arp~299-A.
The noise of this image is 18.5 $\mu$Jy/beam with a peak of 907 $\mu$Jy/beam
and a resolution of $10.0 \times 8.0$ mas, and is the deepest image ever 
of the central 150 pc of Arp 299-A at 5.0 GHz. 
The number of sources above the $5\sigma$ cutoff threshold was 25
  (see Fig.~\ref{fig,mean}). 
All peaks above 5 times
the noise in the average image were identified, and all their locations were
checked in each individual epoch.
All of them had a counterpart above
  $4\sigma$ in at least two out of six epochs, thus confirming them as
  real sources. We note that 20/25 sources were already identified in
  \citet{PT09}, based on two eEVN observing epochs.
To account for the
  possible existence of real, fast-varying sources (e.g. due to
rapidly evolving type Ib/c SNe or microquasars), we also searched for
sources which were fainter than the $5\sigma$ cutoff in the average
image, but had a peak above $7\sigma$ in a single epoch. One source
($33.6211+46.692$, henceforth A27) has been found and included in our
list. The measured peak flux of A27 at experiment RP014C is
267 $\mu$Jy/beam corresponding to a  $7.1\sigma$ detection. At all other epochs
the maximum peak brightness measured in a circular region with a radius of 5 
pixels  centered on the position of A27 is 38 $\mu$Jy/beam (RP009),
57 $\mu$Jy/beam (RP014A), 78 $\mu$Jy/beam (EP063B), 75 $\mu$Jy/beam (EP068A)
and 55 $\mu$Jy/beam (RP014D).
Therefore, the total number of sources in Table~\ref{tab,sources1}
and Table~\ref{tab,sources2} is 26.

We first fitted the sources with Gaussian components using the AIPS
task JMFIT. At the resolution of $10\times 8$ mas, corresponding to a
linear resolution of about 2 pc, all the sources resulted consistent
with being point-like and therefore we preferred to estimate the peak
flux using the task MAXFIT which fits a quadratic function and does
not assume a Gaussian intrinsic shape for the source.  In the
following, we will use the peak flux density derived with MAXFIT as
the total flux density.  Table~\ref{tab,sources1} lists the 26 sources
detected in the Arp~299-A field.  For each source we give the name, RA
and DEC offsets in mas from the pointing position (RA=11:28:33.622,
DEC=$+$58:33:46.610), the flux density and corresponding luminosity in
the average image, the highest flux density and corresponding
luminosity among the six epochs. Finally, the last column lists the
source code as reported in \citet{PT09} if present, or a new
designation (from A26 to A31) for sources previously undetected
at 5.0 GHz.

The measured fluxes at all 6 epochs for the {\it bona-fide} sources
are listed in Table~\ref{tab,sources2}. Flux densities between
$3\sigma$ and $5\sigma$ are given between brackets, while for sources
with fluxes below $3\sigma$ at a given epoch, the upper limits are given.
The last column of this Table is the variability flag (see
Sec.~\ref{sec,var}).  
It is important to note that all fluxes given in Table~\ref{tab,sources1} 
and Table~\ref{tab,sources2} are the measured fluxes, scaled using the
image amplitude correction factor as derived in
Section~\ref{sec,epoch}. 

We note that sources A1 and A5 should not be regarded as possible SNe or
SNRs as they have already been identified as components of a low-luminosity 
AGN \citep{PT10}.

\subsection{Assessing image reliability}
\label{sec,sim}

The Arp~299-A field is challenging for VLBI observations because of
the large number of weak ($\lsim 1$ mJy) radio sources over a rather large
field of view ($\simeq 0.7\times 0.5$ arcsec). High sensitivity observations
with good {\it u-v} coverage are needed to properly recover the
right flux at the right position in the sky. Therefore, the
reliability of the VLBI image, and the parameters that can be derived from it,
is an issue that must be tackled and
assessed.  We did this by using the AIPS task UVMOD to produce
a data set with the same {\it u-v} coverage and noise level as the
real one, but with mock sources with known fluxes and positions. The
parameter FACTOR was set to zero, to keep only the model.  This
allowed us to inject the mock sources at the same position as the real
ones.  The parameter FLUX was tuned (values in the range 1.2 -- 1.5
mJy/weights, depending on the epoch of observation) in order to have
the same scatter in the model data-sets compared to the real ones.
For each epoch, one simulation was done using positions and  fluxes as
given in Table~\ref{tab,sources2}. To increase statistics, 4 more
simulations were done injecting the sources at the same positions, but
allowing the flux to vary up to $\pm 60$ $\mu$Jy with respect to the
value listed in Table~\ref{tab,sources2}. At each epoch, a total of
about 100 mock sources splitted in five simulations was used.  The mock
data sets were then imaged, using the same parameters of the real
data, and the fluxes and positions of the injected sources were
measured.

We splitted the mock sources in three different input flux density bins 
and derived the mean value and dispersion of the difference between  the 
injected and the measured flux density. 
The results are summarized in Table~\ref{tab,mock}.

\begin{table*}
\centering                          
\caption{Simulations}             
\label{tab,mock}      
\renewcommand{\footnoterule}{}
\begin{tabular}{rrrrrrrrrrrrrrrrrr}
\hline \hline
 Input Flux &\multicolumn{2}{c}{RP009}&\multicolumn{2}{c}{RP014A}&
\multicolumn{2}{c}{EP063B}&\multicolumn{2}{c}{RP014C}&
\multicolumn{2}{c}{EP068A}&\multicolumn{2}{c}{RP014D}\\ 
  &Mean$^\mathrm{(a)}$&$\sigma_n^\mathrm{(b)}$&Mean&$\sigma_n$&Mean&$\sigma_n$&Mean&$\sigma_n$&Mean&$\sigma_n$ &Mean& $\sigma_n$ \\
\lbrack$\mu$Jy\rbrack &\multicolumn{2}{c}{\lbrack$\mu$Jy\rbrack} 
&\multicolumn{2}{c}{\lbrack$\mu$Jy\rbrack} &\multicolumn{2}{c}{\lbrack$\mu$Jy\rbrack} 
&\multicolumn{2}{c}{\lbrack$\mu$Jy\rbrack} &\multicolumn{2}{c}{\lbrack$\mu$Jy\rbrack} 
&\multicolumn{2}{c}{\lbrack$\mu$Jy\rbrack} \\
\hline 
150 -- 300  & $-$14 & 30 & 1 & 30 & $-$5 & 22 & 1 & 37 & $-$9 & 32 & 11 & 25\\
300 -- 600  & 9 & 24 & 15 & 30 & $-$8 & 22 & 6 & 34 & $-$1 & 29 & 29 & 26 \\
$> 600$     & $-14$ & 24 & $-$2 & 31 & 7 & 21 & 21 & 35 & $-$11 & 21 &$-$29&
17 \\
\hline
\end{tabular}   
\begin{list}{}{}
\item[$^{\mathrm{(a)}}$] Mean value of the difference between injected and 
recovered flux. 
\item[$^{\mathrm{(b)}}$] Dispersion of the difference between injected and 
recovered flux. 
\end{list}

\end{table*}

We also checked the difference between the injected peak position and
the position of the measured peak. The results were very good,  with
a dispersion of this difference $\lsim$ 1 mas even in the faintest flux
density bin.  The results from these simulations can be summarised as
follows.  At each epoch, the mean value of the flux difference is
consistent with being zero and/or shows both positive and negative
values at different flux bins. This implies there is no significant
systematic offset between injected and measured fluxes. The
dispersion in the flux difference, $\sigma_n$, provides us an
empirical estimate of the epoch and flux density dependent random noise error 
affecting our flux
measurements and it will be used later when testing the variability of
the compact sources in the Arp~299-A field.

\begin{figure*} \centering
 \includegraphics[width=150mm,angle=0]{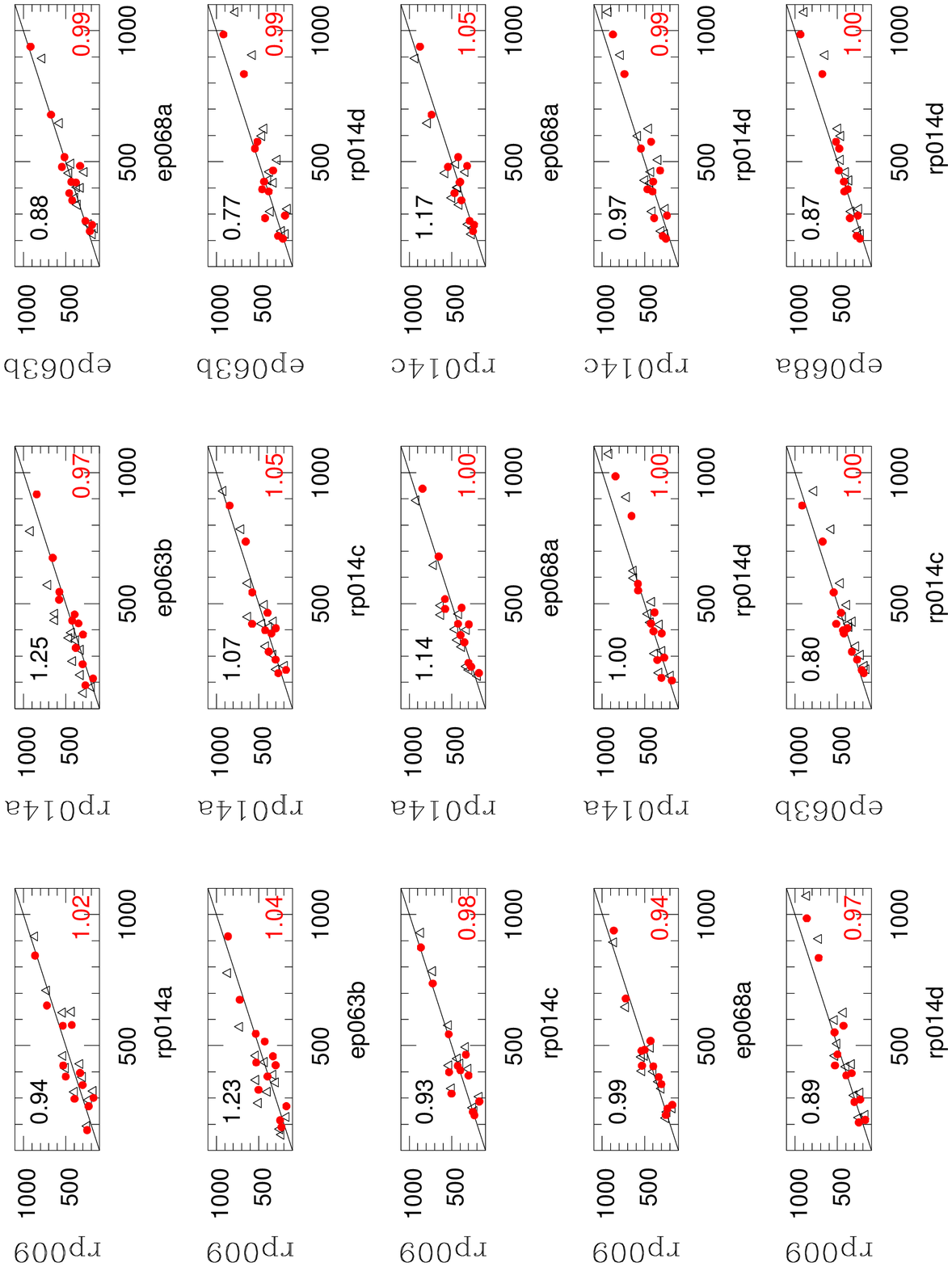} \\
 \caption{{\bf (To be seen in landscape)} Measured flux of the 12 
brightest sources at two different epochs for all the 15 possible combinations.
 Flux units are $\mu$Jy. The solid line is
     the equal flux line. Open triangles are the originally measured fluxes 
 and the number at the top-left corner of each panel is
     the median of the ratios between the two epochs. Filled dots are the
fluxes after the epoch-dependent correction listed in Table~\ref{tab,scale}
are applied, and the number at the bottom-right corner of each panel is the 
median of the ratios between the two epochs. }
 \label{fig,scale1}
\end{figure*}

\subsection{Multi-epoch analysis}
\label{sec,epoch}

From the previous subsection we have derived an estimate of the noise
error, $\sigma_n$, affecting our images and therefore our flux density
measurements.  In this subsection, we take full advantage of all the six
epochs of observation to derive systematic scale offsets in the
amplitudes of our images.  To do so, we identified the 12 brightest
objects detected at all epochs. Eleven of these sources (A0, A1, A2, A3, 
A4, A5, A7, A8, A10, A11, A12) have $\ge 5\sigma$
detections at all epochs with the remaining object (A15) detected with 
SNR $= 4.5$ at one epoch, RP009, which is the observation with the highest
noise.  Then, for each of the 15 possible combinations of two
different epochs, we plotted the measured flux at one epoch
versus the measured flux at the other epoch for the 12 sources
and all epochs.
These ratios are shown in Fig.~\ref{fig,scale1} as open triangles
for all the 15 possible combinations
The number in the top-left corner of each panel in
Fig.~\ref{fig,scale1} is the median of the ratio of the fluxes of the
12 sources.  Clearly, variability adds scatter to the points with
respect to the equal flux line, but we should roughly expect the same
number of sources above and below this line, and therefore median
values close to unity.  This is clearly not the case, meaning that
there are still some calibration issues at some of our epochs.  
These calibration offsets can be produced by errors in 1) the Tsys-based 
a-priori amplitude calibration, 2) the bandpass calibration, and/or
3) the transfer of
amplitude corrections from the phase reference source to the target including
possible coherence losses, as shown for example by \citet{Mart10}.
We applied epoch-dependent scale factors to the fluxes in order to
minimize the deviation of the medians from unity.  The derived scale
factors are listed in Table~\ref{tab,scale} and the corrected fluxes are 
shown in  Fig.~\ref{fig,scale1} as filled dots.
The number in the bottom-right corner of each panel in
Fig.~\ref{fig,scale1} is the median of the ratio of the fluxes of the
12 sources after the correction.
We note that all the 12 sources used to derive the scale factors were 
already detected by Ulvestad, three years earlier than our observations. 
Therefore,
it is highly unlikely that most of these sources are SNe which should display 
a monotonic decrease of the flux density. This is consistent with the fact
that the derived scale factors do not show a monotonic trend with time that 
would signal a coherent decrease (or increase) in the flux densities of the 12 
sources.

\begin{table}
\centering                          
\caption{Image scale factors}             
\label{tab,scale}      
\renewcommand{\footnoterule}{}
\begin{tabular}{llllll}
\hline \hline
 Date & Experiment  & Scale Factor \\
      & Code        &            \\
\hline 
08 Apr 2008 & RP009   & 1.00 \\
05 Dec 2008 &  RP014A & 0.92 \\
12 Jun 2009 & EP063B & 1.18 \\
11 Dec 2009 & RP014C & 0.94  \\
28 May 2010 & EP068A &  1.05 \\
24 Nov 2010  & RP014D & 0.92 \\
\hline
\end{tabular}   
\end{table}

After applying these scale factors, the mean value of the 15 medians 
shown in the bottom-right corner of each panel in Fig.~\ref{fig,scale1} is
 1.000 with a dispersion of 0.029. Therefore, we take 2.9\% as the residual
calibration error, $\sigma_{cal}$, after the scale factors are applied.

\section{Discussion}

\subsection{Comparison with previous observations}
\label{sec,comparison}

Here, we compare our results with those previously found by
\citet[][henceforth PT09]{PT09} and \citet[][henceforth U09]{Ulve09}.
As already stated, PT09 reported on the first two epochs of
observations (experiments RP009 and RP014A).  20/26 objects
  detected in PT09 are confirmed after 6 epochs by our average image,
  and only 6/26 are not confirmed by our new, more robust, selection
  criterion.  These six sources were detected in only one
epoch (RP014A) in PT09, and were among those with the faintest flux
densities. Two of these sources (A16 and A24) have a $\simeq
4.5\sigma$ counterpart in the average image (peak flux of $83~\mu$Jy
for both sources) and future observations might confirm they are real
sources close to the detection threshold.  The remaining four objects
could have been spurious sources detected above the $5\sigma$ threshold
because of a slight underestimate of the noise in the previous
analysis, and are not confirmed by our new selection criteria,
  which are more robust against artifacts. 
On the other hand, one cannot simply rule out the possibility of
sources which are detected in a single epoch, like A27 but at fainter 
flux density.

The comparison with the observations reported by U09 is also very
interesting as those were done in the period around 2004-2005, three
years before our first observations. U09 observed with the VLBA+GBT at
both 2.3~GHz and 8.4~GHz, reaching an r.m.s. of about 20 $\mu$Jy/beam at
both frequencies in his averaged images, which is a value very similar to the 
sensitivity in our averaged 5.0 GHz image. The 8.4 GHz observations are more 
appropriate for
  comparison purposes, since this observing frequency is closer to
  ours (5.0 GHz), and hence less affected by spectral effects, e.g. 
some sources could be affected by free-free absorption at frequencies around 
and below 2.3 GHz due to foreground H~II regions, like in the case of A0 
\citep{PT10}. Twelve out of the 13 objects detected by U09 at 8.4~GHz are 
confirmed above $5\sigma$ and the
remaining one has a $4\sigma$ counterpart in our average image. This
implies that most of these sources are long-lived ($\gsim 6$ years)
objects.

On the other hand, there are eight sources detected in our 5.0 GHz observations 
that are not detected by U09 at neither frequency. 
These are A13, A14, A19, A25, A26, A27, A28 and A29. 
One source (A14) is the furthest from the  image centre and might have not been 
detected by U09 for this reason. The remaining seven sources are among 
the weakest ones, but we note that there are equally weak sources 
(A18, A30 and A31) that were detected by U09. 

While it is possible that image sensitivity and image reconstruction
issues could account for the non-detection of some of the sources in the
VLBA+GBT observations carried out in 2004-2005, we suggest that, for at 
least some of these objects, the most likely reason is the explosion of
recent core-collapse SNe (CCSNe). While this point will be developed in 
detailed in Section~\ref{sec,sn}. We note that sources A13, A19, A25 were 
suggested to be recently exploded SNe, and A14 a SNR by PT09, whereas
sources  A26 to A29 were not detected in those observations.

The 5.0 GHz radio luminosity of the compact objects in Arp~299-A covers
almost an order of magnitude, (3--20)$\times 10^{26}$
erg~s$^{-1}$~Hz$^{-1}$, with a median L$_{\rm 5GHz} = 6.9\times 10^{26}$
erg~s$^{-1}$~Hz$^{-1}$ (we have excluded sources A1 and A5 since these are 
associated to the low luminosity AGN), and is systematically lower when 
compared with the mixed population of SNe and SNRs in Arp220 which have a
 median L$_{\rm 5GHz} = 60\times 10^{26}$ erg~s$^{-1}$~Hz$^{-1}$ \citep{Bate11}.
Unfortunately, luminosity by itself does not allow to discriminate
between SNe and SNRs.  Indeed, CCSNe can have a wide range of peak
radio luminosities ($10^{25} - 10^{29}$ erg~s$^{-1}$~Hz$^{-1}$)
depending on the wind properties \citep[e.g.][]{Albe06,Romero11}.  
Our stacked image is sensitive to objects with a luminosity 
L$_{\rm 5GHz} \geq 2.3 \times 10^{26}$ erg~s$^{-1}$~Hz$^{-1}$ (five
times the r.m.s.), and thus allows the detection of sources with a steady, 
faint radio emission which would otherwise go undetected. 
Those sources must therefore be long-term emitters, i.e., relatively old CCSNe 
and/or SNRs.
Our true sensitivity limit for detecting recently exploded SNe is given by 
the r.m.s. of each single epoch. From Table 4, the average r.m.s. is of 
$\simeq 33\,\mu$Jy/beam, our luminosity sensitivity for a threshold of 
$5\sigma$ is then L$_{\rm 5GHz} \simeq 4.0 \times 10^{26}$ erg~s$^{-1}$~Hz$^{-1}$,
 implying that our observations are sensitive enough to allow the detection of 
at least some Type IIP SNe, which are known to be relatively faint
radio emitters.
We also note here that our 2.5-yr monitoring of Arp 299-A did not reveal
a radio supernova similar, or brighter than A0,  which exploded
back in 2003, and most likely was a Type IIn event.
This is unlike the case of Arp 220, where essentially every new
detected radio supernova shows a radio power which is typical of Type
IIn supernovae. 

The radio emission from SNRs is expected to peak at the
beginning of the Sedov phase \citep{huang94}, and the peak 5.0 GHz SNR
luminosity can be roughly given by L$_{\rm 5GHz}\approx 3.7\times
10^{24}(M_{ej1}/n_{\rm ISM})^{-0.53}$ erg~s$^{-1}$~Hz$^{-1}$
\citep[scaled to 5.0 GHz from][assuming $\alpha=-0.7$]{Parra07}.  where
$M_{ej1}$ is the ejecta mass in units of $10 M_\odot$ and $n_{\rm
  ISM}$ is the interstellar medium (ISM) number density in cm$^{-3}$.  
The molecular number
density in Arp~299-A is $n_{ISM}\sim 10^4$ \citep{Aalto97} and for
$M_{ej1}=0.5$, we obtain a typical SNR 5.0 GHz luminosity of $7.0\times
10^{26}$ erg~s$^{-1}$~Hz$^{-1}$ totally consistent with the median
luminosity of the compact objects in Arp~299-A.  Thus, the
  observed 5.0 GHz radio luminosities of the compact sources in
  Arp~299-A are in agreement with expectations from a population of
  relatively young CCSNe,  SNRs, or a combination of both.

\subsection{The variability and nature of the compact sources in Arp299-A}
\label{sec,var}

Variability can be an important tool to help classifying the compact
sources in Arp~299-A as radio supernovae, supernova remnant, AGN or
other more exotic classes of sources.  For the sources in
Table~\ref{tab,sources2} with flux measurements $\ge 3\sigma$ at all six 
epochs, we calculated the chi-square:

$$\chi^2 =\sum_{i=1}^6 {S_i-\bar{S}\over \sigma_{\epsilon,i}^2}$$

where $\sigma_\epsilon = \sqrt{\sigma_n^2 +\sigma_{cal}^2}$ is the
combination of the noise error (see Sect. \ref{sec,sim}) and the residual
calibration error (see Sect. \ref{sec,epoch}) and $\bar{S}$ is the weighted
average flux over the six epochs.  We considered variable (V) the
sources with a probability that the flux variations were due to random
changes $p(\chi^2) < 1\%$, possibly variable (P) those with $1\% \le
p(\chi^2) < 5\%$ or otherwise non-variable (NV).  The variability code
is listed as last column in Table~\ref{tab,sources2}.  The 19 sources
for which this analysis was possible are evenly distributed in 
three categories: 7, variable, 5 probably variable and 7 not variable.

Figure~\ref{fig,light1} shows the derived light-curves for 5 variable
sources and 5 non-variable sources. The time coverage is limited and
the errors are relatively large, but nevertheless it is clear that
some sources show significant variability and this variability has
different characteristics. In the  2.5 years covered by our
observations, we find sources which display a rather smooth increase
and/or decrease in the flux density on a relatively long timescale
(e.g. A0, A11 and A7 in the first two years) and sources with more
erratic and faster variations (e.g. A2, A12 and the sharp increase in
A7 on May 2010). Variability with similar characteristics has been
observed also in the compact objects in Arp~220 \citep{Rovi05}.

\begin{table}
\centering                          
\caption{Variability and spectral index}             
\label{tab,spix}      
\renewcommand{\footnoterule}{}
\begin{tabular}{llllll}
\hline \hline
 Source & Var.Flag  & $\alpha^\mathrm{(a)}$ \\
      &        &            \\
\hline 
A3 & NV & $-0.46$ \\ 
A4 & NV & $-0.46$ \\
A5 & NV & $<-0.93$ \\
A8 & NV & $<-0.58$ \\
A0 & V  & $> 1.28$ \\
A2 & V  & $-0.21$ \\
A7 & V  & $0.07$ \\
A11& V  & $-0.14$ \\
A12& V  & $-0.22$ \\
\hline
\end{tabular}   
\begin{list}{}{}
\item[$^{\mathrm{(a)}}$] Spectral index calculated between 2.3 and 8.4 GHz
from \citet{Ulve09} with the exception of A0 where $\alpha$ is between 1.7 and 
5.0 GHz from \citet{PT10}.
\end{list}
\end{table}

We also note that 
four sources classified as non-variable (NV) and five classified as variable (V)
have a two-point spectral index measured between 2.3 and 8.4 GHz,  
or an upper/lower limit in U09. The values are listed in 
Table~\ref{tab,spix}, where $\alpha$ is defined as $S_\nu \propto \nu^\alpha$.

Coincidentally, there is a correlation between the radio spectral
index measured between 2.3 and 8.4 GHz observations carried out around
2004-2005 and the flux density variability at 5.0 GHz in the period
2008-2010. The non-variable sources have steep spectral index ($\lsim
-0.5$), while the variable sources have flat, or inverted, spectral
index ($\gsim -0.2$). Source A5 has been already identified with a
bright knot/hot-spot component in the A1-A5 complex by \citet{PT10},
where A1 was identified with the low luminosity AGN in Arp~299-A. The
steep spectrum and lack of variability are consistent with this
hypothesis. 
For the remaining three sources, their steep radio spectra
and lack of significant variability are strong indicators that the
sources are likely SNRs passively evolving in the interstellar medium.

\begin{figure} \centering
 \includegraphics[width=80mm,angle=0]{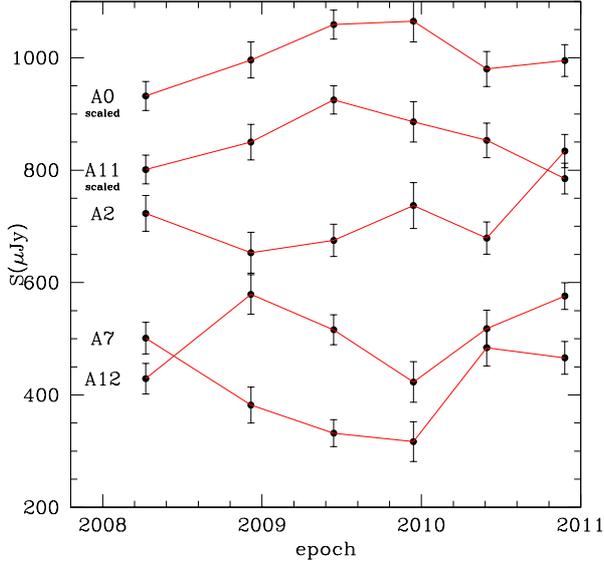} \\
 \includegraphics[width=80mm,angle=0]{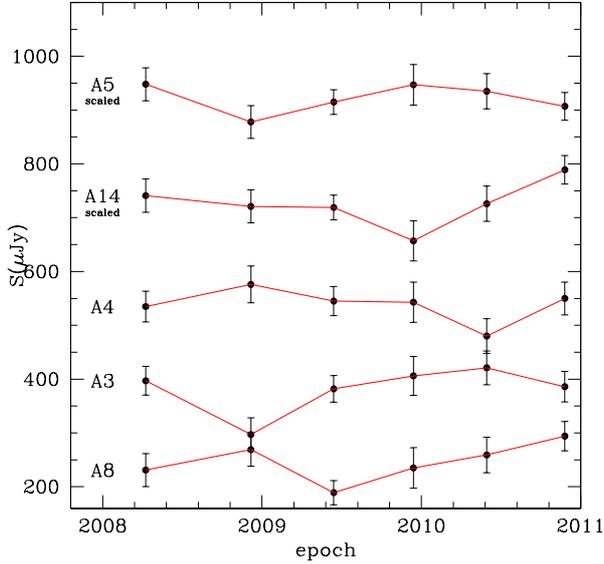} \\
  \caption{{\it Upper panel}: light curves for 5 sources classified as variable.
  The data for source A11 and A0 have been scaled up by 500 $\mu$Jy and 600 
$\mu$Jy respectively to avoid confusion. {\it Lower panel:} light curves for 
5 sources classified as non-variable. The data for A14 and A5 have been scaled 
up by 500  $\mu$Jy and 700 $\mu$Jy respectively to avoid confusion.
} 
 \label{fig,light1}
\end{figure}

On the other hand, all objects showing significant variability for which 
the previous observations allowed to derive a spectral index, have
flat or inverted radio spectra. The identification of the nature of these 
sources is 
more complex. The radio light-curve of SNe is determined by the competition 
between decaying synchrotron emission and decreasing free-free absorption,
producing radio peaks that occur progressively later at longer wavelenghts
\citep{Weil02}. Such a radio light curve evolution yields also to spectral 
indices, which evolve from inverted to flat  --during their optically
  thick phase, while the supernova is its way to the peak flux
  density-- and then become steep, once the supernova is in its
  optically thin phase for all the compared frequencies, 
well past the peak.
None of our sources is neither fading away monotonically, nor rising and 
falling coherently as it would be expected from a SN.
After a rise to a maximum, the flux density of a radio SN generally declines,
although in some case it is known to fluctuate because of inhomogeneities
in the supernova circumstellar environment \citep{Weil02, Mont00}.
All our variable sources show both increase and decrease of the radio 
flux density in the period, about 2.5 years, covered by our monitoring.
Only one of them, A0, has been  clearly identified with a radio SN that reached
its maximum around 2003 \citep{Neff04, Ulve09, PT09}. 
The smooth variability in A0 during the period of
our monitoring could be an example of the fluctuations observed during the
decaying phase in other SNe \citep{Weil02, Mont00}.

Most compact objects in Arp 220 are classified either as SNe or as
SNRs \citep{Parra07}. However, recent multi-wavelength observations
have shown evidence for some objects being in a transition phase, from
SN to SNR \citep{Bate11}. This could easily be the case for some of
the objects in Arp~299-A as well. For example, the source labelled as
A7 (see Fig.\ref{fig,light1}) displays a steady decay of flux density
for at least two years, followed by a fast ($\sim 0.5$ yr) and strong
brightening ($50\%$ of flux density increase) which is suggestive of
A7 experiencing a transition from SN to SNR.  Those 'bursts' in flux
density would be produced by the impact of the freely expanding
forward supernova shock into the ISM.

Our VLBI monitoring at 5.0 GHz is continuing and we are confident that 
the forecoming observations, together with the spectral indices 
for a larger 
number of sources derived from simultaneous 1.4 and 5.0 GHz observations, will 
allow to shed more light on the variability-spectral index relationship 
suggested by Table~\ref{tab,spix}.
 
\subsection{Extended and compact emission in Arp~299-A}
\label{sec,ext}

\begin{figure} \centering
 \includegraphics[width=80mm,angle=-90]{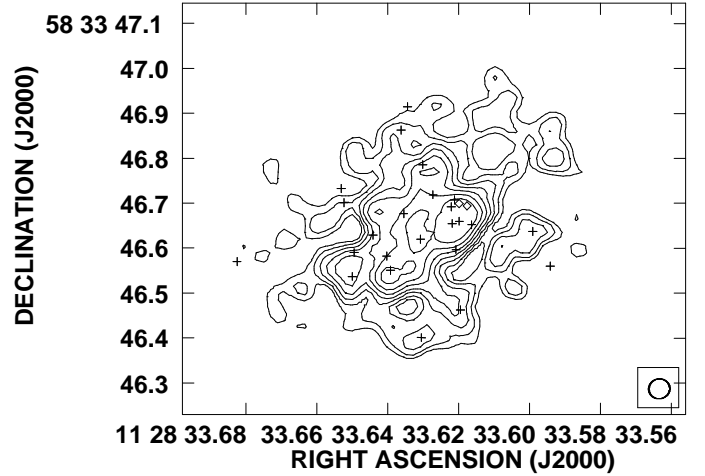} 
\caption{ MERLIN image at 5.0 GHz with  a resolution of
46 $\times$ 44 mas at position angle -75$^\circ$.  The beam is shown at the 
bottom-right corner. Contours levels are a geometric progression in square root of 2, starting from 5 times the off-source r.m.s of 64 $\mu$Jy/beam.
The position of the 26 compact radio sources detected by our EVN observations 
are shown with crosses or diamonds (A1 and A5)} 
 \label{fig,merlin}
\end{figure}

We show in Fig.~\ref{fig,merlin} the contours of the MERLIN image at 5.0 
GHz with a resolution of $46\times 44$ mas. The position of the 26 compact
radio sources detected by our EVN observations are shown with crosses or 
diamonds.
The MERLIN data on Arp 299 were obtained in the course of a
Target of Opportunity observation aimed at searching for radio
emission from supernovae 2010O and 2010P, which resulted in upper
limits to their radio emission \citep{Beswick10}.
The EVN is essentially sensitive to very compact emission, while
MERLIN is also sensitive to the more diffuse emission between the
compact components and has a resolution of $46\times 44$ mas.  
The flux density detected in compact sources above $5\sigma$, as imaged with 
the EVN, is on average $\simeq$ 7.3
mJy. This compact flux is spread over an area of $\approx 0.71 \times
0.68$ arcsec$^2$ 
($\simeq 154 \times 148$ pc$^2$).  The total flux detected above $5\sigma$ in 
the MERLIN image is $\simeq 97\pm 10$ mJy
over an area of $\approx 0.77 \times 0.52$ arcsec$^2$, essentially the
same area covered by the compact sources in our EVN observations.
The MERLIN flux is consistent with the flux ($101.1\pm 5.0$ mJy) measured
by \citet{Neff04} from VLA observations at 5.0 GHz with a resolution of 
$0.38\times 0.31$ arcsec. From \citet{Neff04}, it is possible to
derive a spectral index $\alpha\simeq -0.5$ between 4.9 and 8.4 GHz for a
region fitted with a Gaussian FWHM of $0.38\times 0.30$ arcsec in P.A.
$127^\circ$, which is consistent with the size of the MERLIN detected
extended emission.

The MERLIN emission is dominated by diffuse flux, as 
the EVN-detected compact sources sum up to $< 10\%$ of the total emission
detected in our MERLIN image. 
The radio spectral index of  $\alpha\simeq -0.5$ implies that
the extended radio emission in the MERLIN image  is 
synchrotron emission from relativistic electrons, the bulk of
  which are originally
produced in young SNe and SNRs, since the AGN in Arp 299-A
  contributes very little \citep{PT10} . 
Here, we notice three aspects. First, the compact sources and 
the extended emission trace the same physical region.
As expected, the brightest patch of diffuse emission coincides with 
the region of highest density of compact sources, but some compact sources
are found right at the edges of the diffuse emission as well.
The extended emission has a 5.0 GHz luminosity $L=2.3\times 10^{29}$ 
erg~s$^{-1}$~Hz$^{-1}$.  Since this radio emission must be
  produced by young SNe and SNRs, the inferred 
core-collapse supernova rate, using the standard relation in
\citet{condon92}, is $\nu_{\rm SN} \simeq 0.40 - 0.65$ SN/yr 
for $\alpha$ between  $-0.5$ and $-0.8$. This value is consistent
with the estimate $\nu_{\rm SN} \simeq 0.7$ SN/yr based on the fraction
of far-infrared luminosity attributed to Arp~299-A
\citep{AH00,AH09,Romero11}, and indicates that the main bulk of the
  (recent) star formation in Arp~299-A is most likely happening in its central
  $\sim$150 pc.

 Second, the radiation density in the nuclear region of Arp~299-A
 is so large as to make Inverse Compton (IC) losses 
exceed those by synchrotron radiation.
In fact,  since the energy source, or sources, powering the radio
  emission cannot be larger than the observed radio source size, the
  radiation energy density $u_{\rm r}$ at the surface of a spherical
  radio component covering a solid angle $\Omega$ must be at least
$$ u_{\rm r} \geq \frac{4\, \pi}{c} \left(\frac{\rm
    L_{ir}}{\Omega \, D^2}\right) $$. The solid angle covered
  by the diffuse (MERLIN) radio emission is $\Omega \approx 0.48$
  arcsec$^2$.  
  We first estimated the 1.4 GHz radio emission from the total 5.0 GHz
  flux density measured by MERLIN ($\sim$97 mJy) converted to 1.4 GHz assuming 
  the spectral index, $\alpha \simeq -0.5$. Then,  we obtained the 
  corresponding  FIR luminosity for this region, $L_{\rm IR} \approx 4.8 
\times 10^{10} L_{\odot}$, using the
  standard FIR-to-radio relation (e.g. \citet{yun01}).
  The radiation density is  $ u_{\rm r} \gsim 3.5\EE{-7}$
  \ergcmcmcm.  Since the magnetic energy density is $ B^2 / (8\,\pi)$,
  the magnetic field would have to be at least $B \gsim 3$ mG to
  overcome IC losses. However, the equipartition magnetic field
  corresponding to the diffuse synchrotron radio emission traced by
  MERLIN is $B_{\rm eq} \simeq 160\,\mu$G (here, we assumed a cosmic
  ray protron/electron energy ratio of 100 and unity filling factor),
  which is a very similar value obtained for the central 200 pc of the
  ULIRG IRAS 23365+3604 \citep{Romero12}.
  Therefore, unless strong deviations from equipartition
  exist, IC losses will dominate over synchrotron losses in the
  central $\sim$150 pc-region of Arp~299-A. Thus, the lifetimes of
the relativistic electrons responsible for the 5.0 GHz emission, which
in an equipartition magnetic field would be of $\sim 2.3\EE{5}$ yr,
are heavily cut down to $\lsim$2900 yr due to IC losses.  This
requires that the relativistic electrons responsible for synchrotron
emission in the central $\sim$150 pc of Arp~299-A are being
continuously reaccelerated or, alternativelty, that there is a
continuous injection of fresh relativistic electrons. The natural
places for such acceleration would be the shocks of the SNe and SNRs
that our EVN image reveals as strong synchrotron radio emitters.

 Third, we note that four of the six sources classified 
as non-variable SNe/SNRs (A3, A8, A14 and A29) lie at the outskirts of the 
diffuse, extended emission traced by our MERLIN image in 
Fig.~\ref{fig,merlin} and consequently at the edge of the nuclear starburst 
region. 
This may indicate that all those objects are SNRs, and that the
transition from SN to SNR may happen faster in the outskirst
of the nuclear region of Arp 299-A, where the densities are not 
expected to be so
extreme as in the inner few pc, where intense activity is happening in
scales of a few years (e.g., A0, A27). The non-detection of A0 at low 
frequencies, even several years after its explosion \citep{PT10}, suggests 
that H~II regions (and consequently free-free absorption) are likely more 
frequent in the central region of Arp 299-A, 
which would be in agreement with the above scenario. 
On the contrary, A14 and A29, which were not detected by U09 in his very 
sensitive observations, could be examples of particularly fast evolving 
objects, that would be currently transitioning to SNRs.

\subsection{New CCSNe and the CCSN rate of Arp 299-A}
\label{sec,sn}

\begin{figure} \centering
 \includegraphics[width=80mm,angle=0]{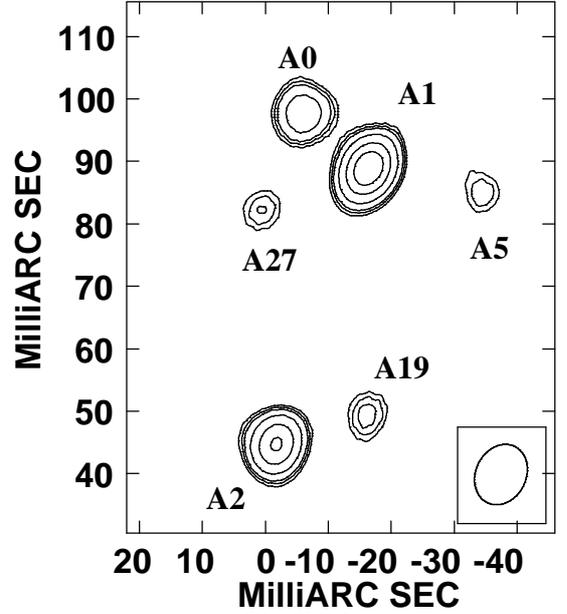} \\
 \caption{Zoom of the central region around component A1 observed 
during experiment RP014C. The beam is shown in the bottom right corner and
the resolution is $10\times 8$ mas.
Contours are drawn at levels -5, 5, 6, 7, 10, 15, 20, 40 times the r.m.s. 
noise listed in Table~\ref{tab,evn6cm}. The compact source A27 is detected 
with $SNR=7$ 
} 
 \label{fig,rp014c}
\end{figure}
As previously mentioned, in the period monitored by our observations none of 
the compact sources in Arp~299-A shows a monotonically decreasing flux 
density which could be interpreted as the optically thin, decaying phase 
of a CCSN. The best chance to identify new SNe is to look for objects without 
a counterpart in the earlier U09 observations. As already said at the end of 
Sec.~\ref{sec,detection}, there are 8 sources detected by the present 
observations at 5.0 GHz, and not detected by U09 at neither 2.3, nor 8.0 GHz.

Based on the far-infrared luminosity 
$L_{\rm FIR} = 2.7\times 10^{11}$ erg~s$^{-1}$ 
for Arp 299-A \citep{CSG02}, and 
assuming the CCSN rate for Arp 299-A is $\nu_{\rm SN}\simeq 0.72$ SN/yr 
\citep{AH00,Romero11} 
the expected number of CCSNe that should have
exploded in the $\sim$5.3 years elapsed between the VLBA+GBT
observations and our latest EVN observation is $\sim$3.7. Since the
explosion of SNe is an intrinsically stochastic phenomenon where
Poisson statistics applies, we should actually expect that the number
of new supernovae be in the range [1.9, 5.6] where the range
corresponds to 1$\sigma$ confidence level \citep{Gehr86}.

Based on the 5.0 GHz radio light-curve behaviour displayed by the
sources above, there seems to be clear evidence for at least two
objects being newly exploded supernovae in the 2.5-yr period covered
by our observations, namely A25 and A27. A25 was already suggested to
be a new SN by PT09, and its behaviour supports a scenario
where it reached its peak luminosity around 2008.9 (+/- 0.3 yr) and
started to slowly decline afterwards. The luminosity at the measured
peak flux density is $6.7\times 10^{26}$ erg~s$^{-1}$~Hz$^{-1}$. Its
behaviour and peak luminosity is consistent with a type II supernova,
most likely a type IIP, or type IIL, based on its relatively modest luminosity
  peak and its slow evolution. 
A27 is detected, with SNR = 7, in just one epoch on December
2009 (see Fig.~\ref{fig,rp014c}), but is undetected at all other
epochs. The corresponding peak luminosity is L$_{\rm 5GHz}= 6.2\times
10^{26}$ erg~s$^{-1}$~Hz$^{-1}$. A search for possible counterparts at
other epochs within  a circular region with a radius of 5 pixels
centered on the
source's peak failed to find emission above a $3\sigma$ threshold. We
have challenged the reliability of A27, imaging the data set with
different resolutions using different weighting schemes and we have
always detected the radio source with an uncertainty on its position
less than 1 pixel (0.5 mas). Since the image size is $2048\times 2048$
pixels the probability of detecting a spurious $7\sigma$ spike of
(Gaussian) noise in the total image area is $< 10^{-5}$. This
probability would be even smaller, if we considered only the
restricted area covered by the detected sources. Therefore, even if
detected at just one single epoch, we are confident that A27 is a real
source. Its flux density rise and decay are fast. At the previous and 
following epochs, 6 months earlier and later, we have upper limits of 78 
$\mu$Jy and 75 $\mu$Jy, respectively. Such a fast rise and decay
of the flux strongly suggests that A27 was a Type Ib/c supernova.

 Since each of our individual observations has a typical r.m.s
  of $\gsim 30 \mu$Jy, which corresponds to a 5.0 GHz luminosity
  $\simeq 1.0\EE{26}$\ergs, we are sensitive to modest radio
  emitting supernovae, including Type IIP SNe, but not
  to very faint radio supernovae. At any rate, our firm discovery of at
  least two CCSNe in our 2.5 yr monitoring corresponds to a lower limit of 0.8
  SN/yr for the CCSN rate in the nuclear starburst in Arp 299-A. 
  Taking into account Poisson statistics, our measured value of the
  CCSN rate, with errors bars corresponding to 1$\sigma$
  errors, is  $\nu_{\rm SN}\gsim 0.80^{+1.06}_{-0.52}$ SN/yr.
This is the first {\it direct} determination of the
  core-collapse supernova rate in Arp~299-A ever, and while it agrees
  with previous estimates, it represents a lower limit to the
  true CCSN rate of Arp~299-A. A more accurate and reliable
  determination of the CCSN rate for Arp~299-A will necessarily
  require a longer monitoring to overcome Poisson statistics.

We emphasize the fact that this value of  $\nu_{\rm SN} $ is a lower
limit to the {\it true}
core-collapse supernova rate of the Arp 299-A system.
Indeed, if we take into account SN2010O, which exploded in Arp
299-A (although far from its nuclear region, at a distance of
$\simeq$1.4 kpc), during the time covered by our obsevations, the overall CCSN rate
for the galaxy would then be  $\nu_{\rm SN}\simeq
1.20^{+0.77}_{-0.43}$ SN/yr. If confirmed by future observations, this value would imply a
larger CCSN rate than expected from empirical conversions of
far-infrared luminosity into CCSN rates.

We also remark here that A27 exploded at a mere 7.2 pc from A0,
the only other source that has been unambiguously identified with a
SN, and at 7.9 pc from the AGN in Arp 299-A. The detection of two
CCSNe in less than 10 years exploding at a (projected) distance of
$\lsim$7 pc apart could suggest that A0 and A27 were part of a 
super-star cluster (SSC), hosting large numbers of massive stars in the 
central region of Arp~299-A. 
SSCs seen in the optical and near-infrared in other
merging galaxies have typical radii $\sim 4$ pc but, in particular
young SSCs (ages less than 10 Myr), may be as large as tens of parsecs
\citep{Whit99}. Near-infrared \cite{Lai99} and optical \citep{AH00}
observations have indeed detected young SSCs in the central regions of
Arp~299. 
A0 was a very bright, long-lasting radio supernova, implying it must have
  probably been a Type IIn SN. A27 was also relatively brigth, but its
  fast evolution indicates that it is most likely a Type Ib/c event. Both A0 and A27 must
  then have had massive  ($M \simeq 18-35 M_\odot$)  progenitors, whose
  lifetimes  are $\lsim$ 10 Myr.  
  Evolutionary models for the
radio emission in starbursts by \citet{perezolea95} show that the SN
rate in an instantaneous burst (for a Salpeter IMF with lower and
upper limits of 0.85 and 120 \msun, respectively) are in the range
(6-11)$\EE{-10}$ yr$^{-1}$\msun$^{-1}$ for a cluster in the time
interval between 3-10 Myr. 
  Since SSCs in galaxies outside the local group and with ages $\lsim$
  10 Myr have at most $M \sim 5\EE{6}$ \msun \citep{portegies10}, 
  the CCSN rate for a SSC of $5\EE{6}$ \msun would then be $\sim 0.0030
  - 0.0055$ yr$^{-1}$, or at most 1 SN every 181 yr. The
  (Poisson) probability of having more than 1 CCSN in such a SSC after
  just $\sim 6.5$ yr is $6.2\EE{-4}$. We therefore rule out the
  possibility that both A0 and A27 exploded in the same SSC, and
  suggest instead that these CCSNe exploded in two different
  SSCs of smaller sizes. The two SSCs could be intrinsically close
or appear nearby due to projection effects.

In addition, four other sources (A13, A19, A26, A29) have radio light curves 
in agreement with them being slowly evolving SNe, similarly to the case
of A0, but the weakness of these sources and the limited number of epochs
prevents a definite conclusion. 
We defer the classification of these objects to another paper, where
spectral information and additional new observing epochs would allow
us to type them more securely. Yet, we stress here that our main
  result (the discovery of at least two new SNe) implies a CCSN rate
  for Arp~299-A of $\nu_{\rm SN}\gsim 0.80$ SN/yr, and 
suggests that most star formation is currently occurring in
the central $\simeq$150 pc of Arp~299-A.

\subsection{Other transient sources}
\label{sec,transients}

In Table~\ref{tab,sources2}, we reported on the variability of the
compact sources in Arp 299-A based on a $\chi$-squared test. This test
was applied to those sources�
whose peak of brightness was above 3 times the noise at all epochs.
Such a requirement was adopted in order to maximize the number of points
in the radio light-curves and therefore better constrain the variability
behaviour, but it fails to account for the possible existence of fast-varying 
objects (e.g., rapidly evolving type Ib/c SNe or other transient sources)
that are not detected at one or more epochs.

Here, we discuss the nature of some of the sources that, while not
fulfilling the criterium for the variability analysis, show significant 
variability albeit of a less extent.
In particular, PT09 already discussed the possibility that source A6
was a microquasar \citep[e.g.][]{Mira94}, 
based on its radio luminosity, radio behaviour at
5.0 GHz, and proximity to two X-ray sources. Now, from our six epochs
of observation, and from its radio longevity and spectrum (it was
already reported at 2.3 Ghz by U09, but not at 8.4 GHz, implying
$\alpha \lsim -1.03$ between 2.3 and 8.4 GHz), we suggest that A6 is
a microquasar which flashes on and off.
A similar behaviour to that of A6 is displayed by A28. In fact, the
source seems to show epoch-to-epoch (i.e., approximately every
six-months) flux density variations of $\sim$40\%, or even higher.
Since A28 was not detected by U09 in his 2004-2005 observations, this
implies that it is a young source ($t \lsim 7$ yr). While it is
possible that A28 could be also another microquasar, we cannot
exclude that it is a different object, even a radio supernovae since,
as we mentioned above, non-standard SNe seem to be relatively normal
in starburst galaxies \citep{Rovi05}.

Nonetheless, we defer a definitive classification of those sources to
a separate publication, where contemporaneous spectral information
will be discussed. 

\section{Summary}
We have monitored the nearby LIRG Arp~299-A at 5.0 GHz for 6 epochs
with the European VLBI Network. The observations have been carried out once 
every 6 months for a period of 2.5 years. The main results of these observations
are as follows.

\begin{enumerate}

\item Twenty-five compact sources are detected above the $5\sigma$
  detection threshold in the 6-epoch average image. One more source,
  even if below the detection threshold in the average image is
  detected with high confidence (SNR=7) at a single epoch and is
  considered as a real fast varying source, probably a type Ib/c SN.
  Two sources (A1 and A5) have been already identified as components of a 
  low-luminosity AGN in Arp~299-A \citep{PT10}.
  The 26 detections occupy a region with a diameter of about 150 pc.

\item Comparison with previous observations shows that most of the compact
sources have been already detected at 2.3 and/or 8.4 GHz in earlier VLBA+GBT
observations with comparable sensitivity \citep{Ulve09}. Only eight sources 
appear to be objects  not reported in U09. 
Some of these new detections
might be due to a slightly larger area imaged by our observations, but we are 
confident that some of them are recently exploded CCSNe.

\item The 5.0 GHz radio luminosities of the compact objects in Arp~299-A
are in the range $3- 20\times 10^{26}$ erg~s$^{-1}$~Hz$^{-1}$ with a median 
of $6.9\times 10^{26}$ erg~s$^{-1}$~Hz$^{-1}$ (A1 and A5 have been excluded since
part of a low-luminosity AGN) about an order of magnitude lower than the 
median luminosity of the population of SNe/SNRs found in Arp~220 
\citep{Bate11}, which suggests
that the population of exploding SNe in Arp~299-A might be
  characterized by a different IMF.  
 A longer monitoring of the compact sources in the
  nuclear region of Arp 299-A should allow us to unambiguously answer
  this relevant question. 
Since the observed radio luminosities are also consistent with typical
SNR luminosities at the beginning of the Sedov phase, we conclude
that the 5.0 GHz observed luminosity are in agreement with expectations
from a population of relatively young CCSNe, or SNRs, or a combination of both.

\item
We have quantified the variability over the 6 epochs (2.5 years)
for the 19 sources  that were detected at all epochs. The sources are
evenly distributed in the variable, probably variable and non-variable classes.
Variable sources can show complex  radio light curves which are difficult to 
interpret. Using spectral index information from the literature we find
that sources classified as variable have flatter radio spectra ($\alpha \gsim
-0.2$) than non-variable sources ($\alpha \lsim -0.5$).
The non-variable and steep spectrum sources could be associated to SNRs
passively evolving  in the ISM. More complex is the identification of the 
variable flat-spectrum sources which might contain SNe and objects in a
transition phase from SN to SNR. A longer monitoring to better characterize
the variability properties together with radio spectral index derived from
simultaneous observations (P\'erez-Torres et al. in preparation) are necessary
to confirm these hypothesis.

\item
A MERLIN image at 5.0 GHz with a resolution of $46\times 44$ mas shows
diffuse, extended emission cospatial with the distribution of the compact 
sources. The extended emission is non-thermal ($\alpha\simeq -0.5$) and can 
be explained with synchrotron emission from relativistic electrons originally 
produced in SNRs. The luminosity of the extended emission implies a 
supernovae rate  of $\simeq 0.40 - 0.65$ SN/yr, which is
consistent with that derived from the far-infrared luminosity of
Arp~299-A, and also in marginal agreement with our direct
  estimate from the new discovered SNe.

\item
We find clear evidence for two recently exploded CCSNe in Arp~299-A.
The object labelled as A25 has reached a maximum around epoch 2008.9 ($\pm 0.3$
yr) and started to slowly decline afterwards. The peak luminosity and the 
relatively slow declining phase suggest that A25 is most likely a
  type II CCSN.
The object labelled as A27 was detected with relatively high confidence 
(250 $\mu$Jy with SNR=7) at a single epoch. Six months later we have an upper 
limit of 75 $\mu$Jy. Since the probability of a fake 7$\sigma$ detection
over the image area is $< 10^{-5}$, we interpret A27 as a fast evolving
type Ib/c supernova. From those two CCSNe, we
obtain a CCSN rate for the innermost $\sim$150 pc of
  Arp~299-A of $\nu_{\rm SN}\simeq$0.80 SN/yr,
  which is in broad agreement with previous estimates at much poorer
  resolution and which is also supported by our CCSN estimate obtained
  from the diffuse, non-thermal radio emission traced by our MERLIN
  observations. This result strongly hints at the bulk of the current star 
  formation taking place in the central $\sim$150 pc of Arp~299-A.

This value represents also a lower limit to the {\it true}
core-collapse supernova rate of the whole Arp 299-A system.
If we consider also SN2010O, which exploded also  in Arp
299-A (although at a distance of $\simeq$1.4 kpc from the nuclear
region) the overall CCSN rate for the Arp 299-A galaxy is  $\nu_{\rm SN}\simeq
1.20^{+0.77}_{-0.43}$ SN/yr, which would imply a
larger CCSN rate than expected from empirical conversions of
far-infrared luminosity into CCSN rates.

\item
Finally, we note that a few objects that were excluded from the variability
analysis because they were undetected ($<3\sigma$) in at least one epoch
show peculiar epoch-to-epoch flux density variations of up to 40\% or larger
(e.g. A6 and A28). These radio sources could be associated to microquasars or
peculiar SNe, and further multifrequency observations are necessary for
a proper classification.

\end{enumerate}

\begin{acknowledgements}
  The continuing development of e-VLBI within the EVN is made possible
  via the EXPReS project funded by the EC FP6 IST Integrated
  infrastructure initiative contract \#026642 - with a goal to achieve
  1 Gbit/s e-VLBI real time data transfer and correlation.  The EVN is
  a joint facility of European, Chinese, South African and other radio
  astronomy institutes funded by their national research councils.
  MAPT, RHI, and AA acknowledge support by the Spanish MICINN
  through grant AYA 2009-13036-C02-01, cofunded with FEDER funds.
 The authors thank an anonymous referee for providing us constructive
 comments and suggestions.
\end{acknowledgements}


\Online

\end{document}